%% file: main.tex
\documentclass{article}

\usepackage{PRIMEarxiv}

\usepackage[utf8]{inputenc} %
\usepackage[T1]{fontenc}    %
\usepackage{subfig}
\input{my_packages}

\title{Structure-Preserving Graph Contrastive Learning for Mathematical Information Retrieval
}

\author{
  Chun-Hsi Ku, Hung-Hsuan Chen \\
  Computer Science and Information Engineering \\
  National Central University \\
  Taoyuan, Taiwan \\
  \texttt{jimmyku123@gmail.com, hhchen1105@acm.org}
}

\begin{document}
\maketitle

\input{contents/abs}


\input{contents/intro}

\input{contents/rel-work}
\input{contents/method}

\input{contents/exp}
\input{contents/disc}

\section*{Acknowledgments and GenAI Usage Disclosure}
We acknowledge support from the National Science and Technology Council of Taiwan under grant number 113-2221-E-008-100-MY3. 
The authors used Gemini to improve language and readability. The authors reviewed and edited the content as needed and take full responsibility for the content.

\bibliographystyle{unsrt}  
\bibliography{ref}

\end{document}

%% file: my_packages.tex
\usepackage{color}
\usepackage{booktabs}
\usepackage{multirow}
\usepackage{mathtools}
\usepackage{url}

\newcommand{\figwidth}[4]{
    \begin{figure}[tb]\centering
    \includegraphics[width=#4]{./fig/#1}
    \caption{#2}
    \label{#3}\end{figure}
    {}}

%% file: contents/abs.tex
\begin{abstract}
This paper introduces Variable Substitution as a domain-specific graph augmentation technique for graph contrastive learning (GCL) in the context of searching for mathematical formulas. 
Standard GCL augmentation techniques often distort the semantic meaning of mathematical formulas, particularly for small and highly structured graphs. Variable Substitution, on the other hand, preserves the core algebraic relationships and formula structure.
To demonstrate the effectiveness of our technique, we apply it to a classic GCL-based retrieval model. Experiments show that this straightforward approach significantly improves retrieval performance compared to generic augmentation strategies. We release the code on GitHub.\footnote{ \url{https://github.com/lazywulf/formula_ret_aug}}.
\end{abstract}

%% file: contents/intro.tex
\section{Introduction}

Mathematical Information Retrieval (MIR) is a critical subfield of information science, essential to advance scientific discovery by enabling the effective search and retrieval of mathematical content from vast digital corpora~\cite{sparck1972statistical, wu2015citeseerx}. Effective MIR systems are foundational to next-generation scholarly information access platforms, enabling researchers to navigate the enormous scientific corpus beyond simple keyword matching. Unlike traditional Information Retrieval (IR), which primarily focuses on text-based queries and documents, MIR must contend with the unique structural and semantic complexities inherent in mathematical formulas. Although established IR systems successfully leverage techniques such as TF-IDF and various measures of semantic similarity to retrieve relevant textual documents~\cite{sparck1972statistical, wu2015citeseerx, caragea2014citeseer, chen2013csseer}, these methods often fall short when applied to mathematical content~\cite{mansouri2019tangent}. MIR necessitates models capable of interpreting the intricate syntactic structure of mathematical expressions, recognizing that formulas with different surface appearances can represent the same underlying concept. This inherent characteristic distinguishes MIR significantly from conventional text-centric IR paradigms.

Recent progress in the MIR domain has been driven substantially by the application of sophisticated machine learning techniques, particularly graph neural networks (GNNs) and related methods~\cite {wang24the, pfahler2022self, peng2021image}. These approaches aim to generate meaningful and effective representations (embeddings) of mathematical formulas by meticulously capturing their structure and the complex relationships among constituent mathematical symbols. Previous research has investigated various graph-based and text-based strategies to navigate the specific challenges posed by MIR~\cite{mansouri2019tangent, peng2021image, pfahler2022self, wang24the}. Among these, graph contrast learning (GCL) has emerged as a particularly promising direction, especially for mitigating the common scarcity of labeled relevance data (i.e., explicit relevance scores) in formula retrieval tasks~\cite{wang24the}. GCL frameworks learn robust formula embeddings by treating retrieval as a contrastive problem, aiming to maximize the similarity between different augmented views of the same formula while simultaneously minimizing similarity with unrelated formulas. Models such as MathBERT~\cite{peng2021mathbert} and TangentCFT~\cite{mansouri2019tangent} have demonstrated the potential to integrate both formula structure and the surrounding textual context, with TangentCFT frequently serving as a formidable baseline, especially for retrieval models that rely solely on formula structure.

However, a significant hurdle arises when applying standard GCL methodologies to MIR, specifically regarding the data augmentation step, which is crucial for contrastive learning. The widely used graph augmentation techniques prevalent in the GCL literature, including node drop, edge masking, and feature masking~\cite{you2020graph}, often prove detrimental when applied to the typically small graph structures representing mathematical formulas. Within MIR, the compact nature of formula graphs means that even seemingly minor alterations introduced by these conventional augmentations can drastically distort the formula's fundamental meaning or structural integrity. Removing a single critical operator node or masking an edge signifying a key dependency can easily render the formula syntactically incorrect or semantically nonsensical. This sensitivity arises because nearly every node and edge in a formula graph carries substantial semantic weight. Consequently, employing inappropriate augmentation methods can severely disrupt the vital relationships among mathematical symbols, ultimately impeding the model's ability to learn effective representations and leading to suboptimal retrieval performance.

To address the inherent limitations of conventional augmentations within the MIR context, we introduce a straightforward yet highly effective graph augmentation method tailored explicitly for mathematical formulas, termed Variable Substitution. This technique is designed to introduce the necessary representational variance required for effective contrastive learning while rigorously preserving the core structural and semantic integrity of the original mathematical expression. By strategically focusing on the substitution of variables—elements whose specific identity often matters less than their role within the structure, rather than altering the graph's fundamental topology or critical operator nodes, Variable Substitution effectively navigates the pitfalls of standard techniques. This approach aims to preserve the essential mathematical relationships encoded in the graph structure, thereby addressing the identified shortcomings of existing augmentations for formula graphs.

This paper contributes the following advancements to the field of Mathematical Information Retrieval. First, we introduce Variable Substitution, a simple yet powerful graph augmentation method designed explicitly for MIR, which preserves the essential formula structure during the data augmentation phase of contrastive learning. Second, we present comprehensive experiments demonstrating that Variable Substitution yields significant improvements in formula retrieval performance compared to both existing standard graph augmentation techniques and the established state-of-the-art baseline, TangentCFT~\cite{mansouri2019tangent}. Third, we analyze the efficacy of Variable Substitution across distinct mathematical graph representations, namely Symbol Layout Trees (SLTs) and Operator Trees (OPTs), showcasing its robustness and adaptability by consistently outperforming baseline methods on both structures.

%% file: contents/rel-work.tex
\section{Related Work}

Mathematical Information Retrieval (MIR) poses unique challenges because it requires understanding both the structure and semantics of mathematical expressions. Previous works have introduced various graph-based and text-based approaches~\cite{mansouri2019tangent, peng2021mathbert, wang24the}. Among these, GCL has demonstrated effectiveness in learning formula embeddings by treating formula retrieval as a contrastive learning problem. Models like TangentCFT~\cite{mansouri2019tangent} and MathBERT~\cite{peng2021mathbert} have incorporated both formula structure and text, with TangentCFT serving as a strong baseline for formula-only retrieval models.

Several augmentation techniques have been proposed for GCL, including node dropping, edge masking, and feature masking. However, these approaches often struggle with the small size of formula graphs, where even minor augmentations can significantly alter the formula's meaning. To address this, we propose an augmentation method explicitly tailored to mathematical formulas.

%% file: contents/method.tex
\figwidth{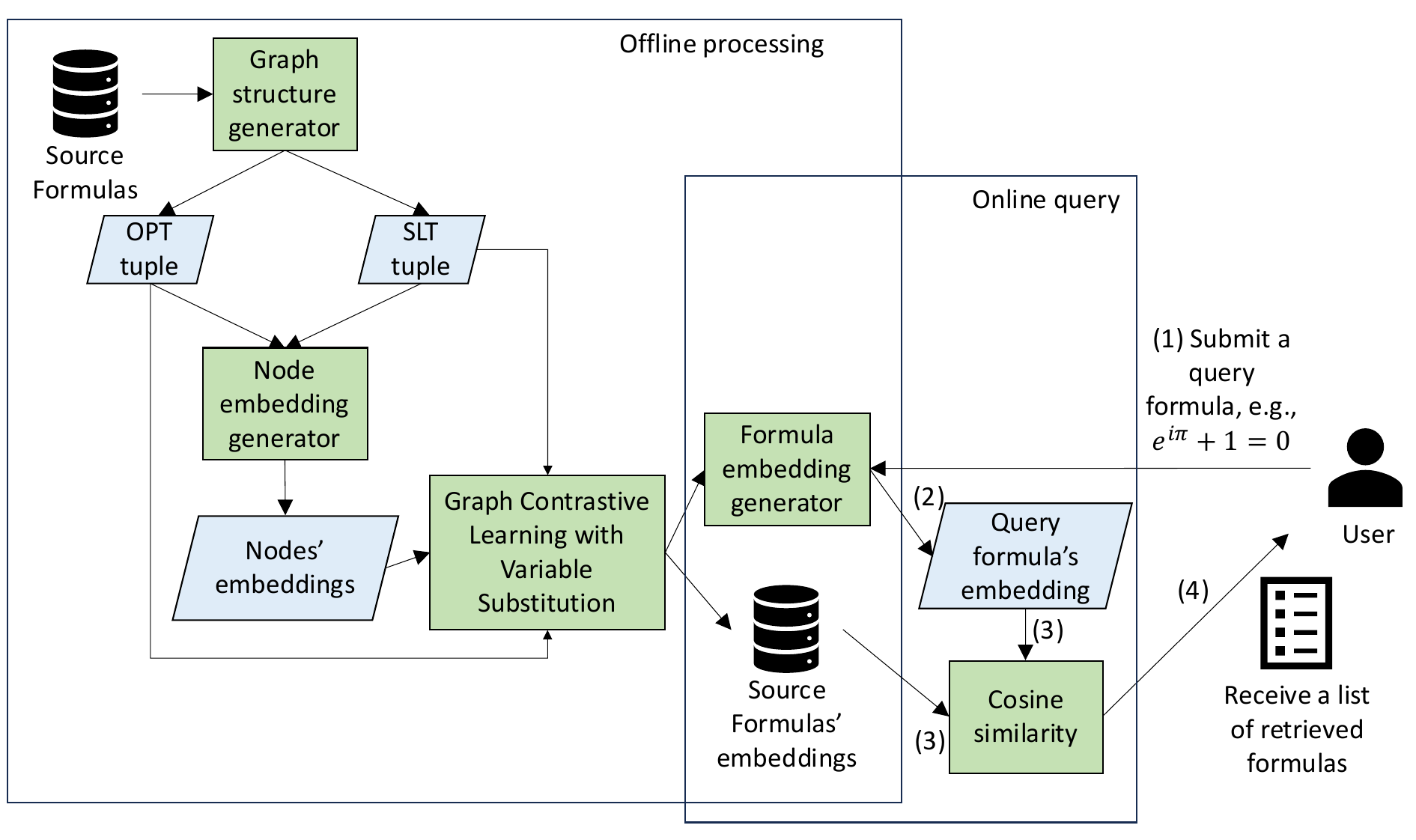}{The online and offline processing of the entire framework; we focus on Graph Contrastive Learning with Variable Substitution in this paper.}{fig:flow}{0.9\textwidth}

\section{Method}

This section outlines our methodology, including graph structure generation, token embedding generation, graph contrastive learning with Variable Substitution, and the online query module. An overview is given in Figure~\ref{fig:flow}.

\subsection{Graph Structure Generator}

Given a mathematical formula, the graph structure generator converts it into graphs that capture the semantic and syntactic relationships among numbers, variables, and operators. We employ two graph structures: the Symbol Layout Tree (SLT) and the Operator Tree (OPT). SLT captures the spatial arrangement of symbols, whereas OPT focuses on operational semantics by representing operators as internal nodes and operands as child nodes~\cite{mansouri2019tangent, wang24the}. These graphs serve as input for the subsequent learning modules.

\subsection{Token Embedding Generator}

The token embedding generator (TEG) utilizes the fastText model to generate embeddings for each node in the graph~\cite{mansouri2019tangent, wang24the, bojanowski2017enriching, joulin2016fasttext}. First, the TEG applies random walks to sample paths from the SLT or OPT graphs. These paths are then encoded using fastText, producing 100-dimensional embeddings for each node. Each embedding reflects the local neighborhood of the symbol within the graph structure, capturing both positional and contextual information. These embeddings serve as the basis for constructing formula-level graph representations.

\subsection{Graph Contrastive Learning with Variable Substitution}

Researchers have increasingly used GCL to generate formula embeddings without relying on labeled relevance scores~\cite{wang24the, you2020graph}. However, popular graph augmentation techniques, such as node/edge dropping or attribute masking, are ill-suited for mathematical graphs. Even minor modifications---like dropping an operator or variable---can fundamentally change a formula's interpretation. Such augmentations are likely to introduce destructive noise, hindering the model's ability to learn meaningful representations.

To address this, we propose a controlled augmentation method, Variable Substitution, which preserves the structural and semantic integrity of formulas. In our GCL setup, we first create an ``augmented view'' of a formula graph by applying Variable Substitution: nodes representing variables are randomly substituted with other variables, and nodes representing numbers are swapped with different numbers. This process alters node identities while preserving the graph's topology. Positive pairs are then formed by the original formula graph and its augmented view. Negative pairs consist of the original graph and any other formula graph within the same training batch. The model is trained to minimize the distance between positive pairs and maximize the distance between negative pairs in the embedding space, thereby learning robust representations that capture the similarity among abstract formulas.

Once the model has learned to generate formula embeddings through contrastive learning, we store them in a database for efficient retrieval. Overfitting is unlikely to be a concern because contrastive learning emphasizes distinguishing between formulas based on inherent structural similarities rather than relying on human-labeled pairs. This enables our model to generalize more effectively to new, unseen formulas, without being constrained by the biases or limitations inherent in human labels.

\subsection{Online Query Module}

The online query module retrieves relevant formulas in response to user queries. When a user submits a query formula, the system generates an embedding for the query based on the trained formula embedding generator. The system then computes the cosine similarity between the query formula embedding and the embeddings of all formulas in the database. Based on these similarity scores, the system ranks the formulas in descending order and returns the most relevant results to the user.

%% file: contents/exp.tex
\section{Experiments}

\subsection{NTCIR-12 MathIR Dataset}
We evaluate our method using the NTCIR-12 MathIR dataset~\cite{kato16report}, a benchmark commonly used for mathematical information retrieval (MIR) tasks. The dataset comprises an extensive collection of mathematical formulas extracted from Wikipedia and relevance judgments for a set of query formulas. The relevance scores are integers between 0 and 4, with higher scores indicating a closer match between the query and the retrieved formulas. This dataset is specifically designed to test both exact and approximate formula-matching capabilities.

\subsection{Evaluation Metrics: Bpref and Full vs. Partial Match}
For evaluation, we use the binary preference metric, bpref, which is particularly suitable for scenarios with incomplete relevance judgments. Bpref measures how often relevant documents are ranked higher than irrelevant ones without assuming that all relevant documents have been labeled in the dataset. This makes it an ideal metric for our experiments, where relevance judgments are limited to a subset of formula pairs.

Since the bpref metric operates in a binary setting (relevant or irrelevant), but the dataset annotations range from 0 to 4, we must apply a threshold to convert the dataset labels to binary. We use two thresholds for this purpose. First, we consider only formulas with a score of 3 or higher to be relevant, and all others to be irrelevant; we refer to this approach as ``full relevance''. Second, we treat only formulas with a score of 0 as irrelevant, and all other scores are considered relevant; we refer to this approach as ``partial relevance''.

\subsection{Compared methods}

We compare Variable Substitution with several generic graph augmentation strategies. To ensure a fair comparison, we use TangentCFT~\cite{mansouri2019tangent} as the base model for all augmentation methods. The compared augmentation strategies include: Node Drop, which randomly removes nodes from the graph; Edge Drop, which randomly removes edges; Node Feature Mask, which masks the features of sampled nodes; and Edge Feature Mask, which masks the features of sampled edges. Finally, the Random strategy randomly selects one of the four aforementioned techniques for each graph.

Since contrastive learning is often sensitive to batch size, we evaluated different batch sizes across all graph augmentation strategies.

\subsection{Results}

\begin{figure}[tbh]
\centering
\subfloat[The full relevance setting\label{fig:slt-full}]{%
\includegraphics[width=0.45\textwidth]{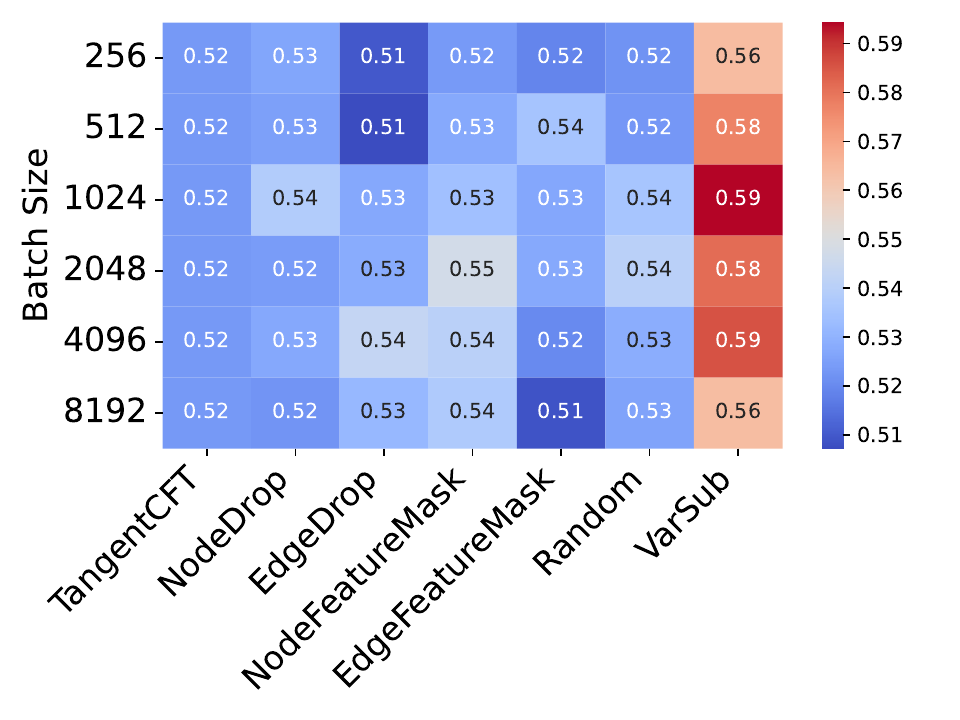}%
}\hfil
\subfloat[The partial relevance setting\label{fig:slt-partial}]{%
\includegraphics[width=0.45\textwidth]{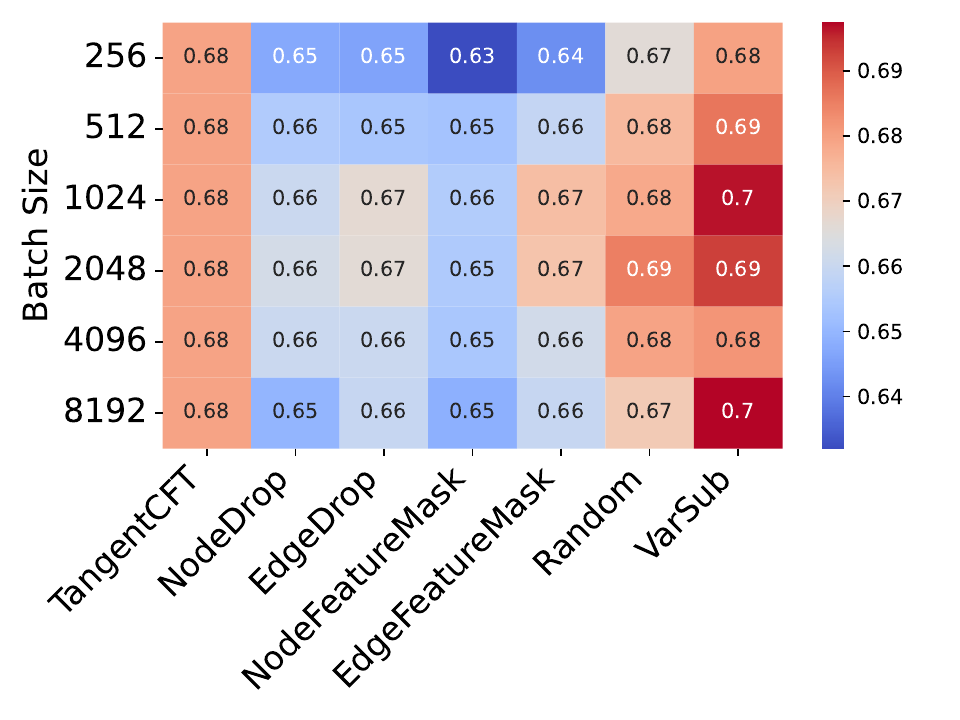}%
}
\caption{The bpref scores using the SLT layout}
\label{fig:slt}
\end{figure}

\begin{figure}[tbh]
\centering
\subfloat[The full relevance setting\label{fig:opt-full}]{%
\includegraphics[width=0.45\textwidth]{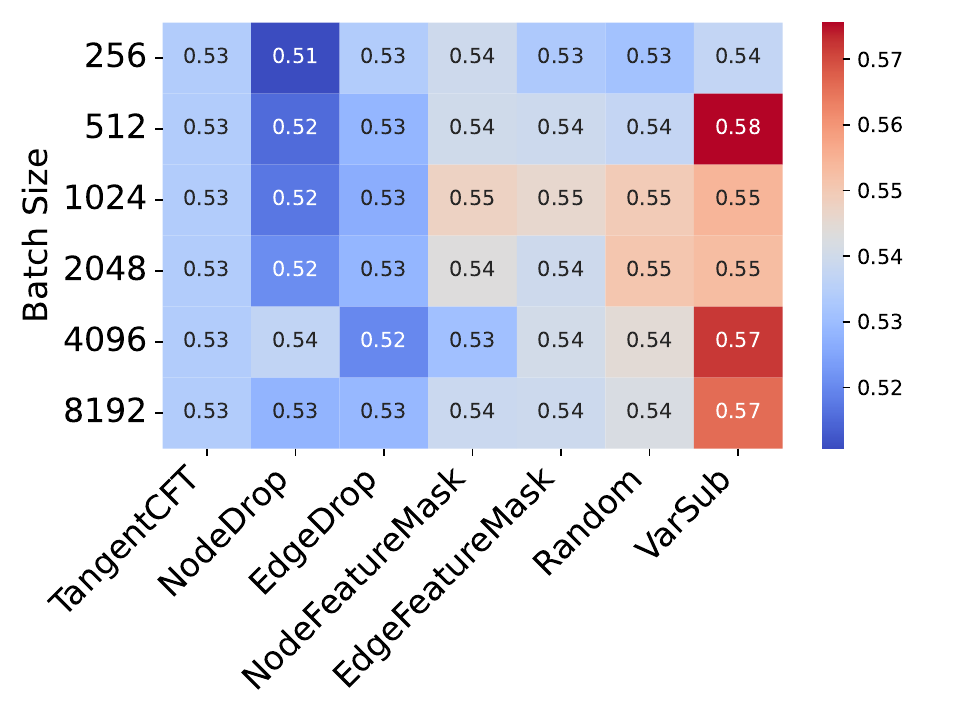}%
}\hfil
\subfloat[The partial relevance setting\label{fig:opt-partial}]{%
\includegraphics[width=0.45\textwidth]{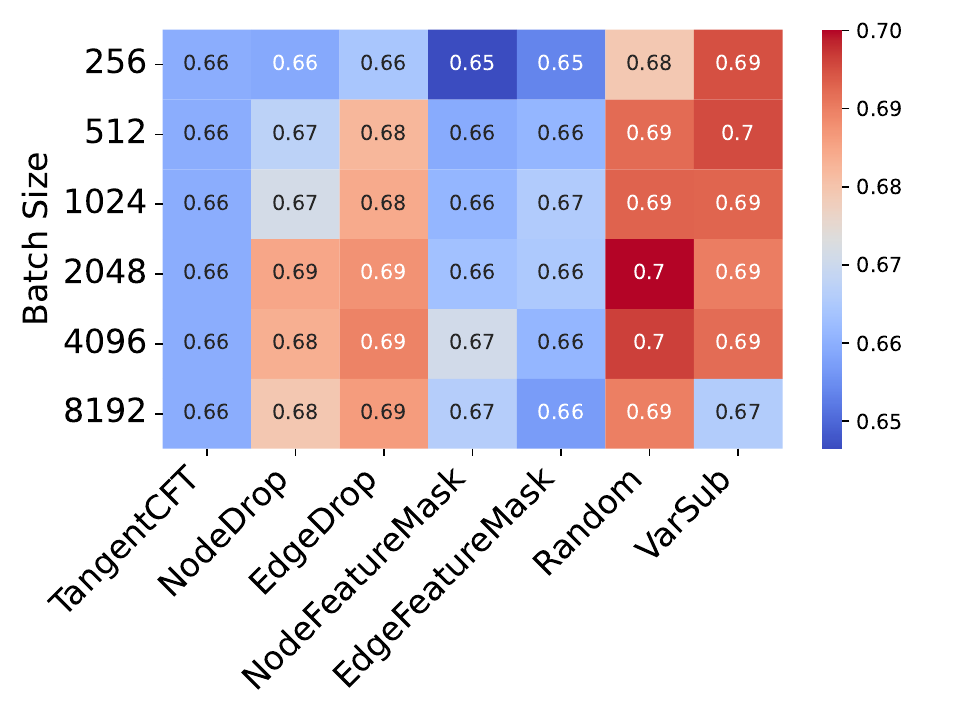}%
}
\caption{The bpref scores using the OPT layout}
\label{fig:opt}
\end{figure}

The results, presented as heat maps in Figure~\ref{fig:slt} and Figure~\ref{fig:opt}, show the performance of different augmentation methods in various batch sizes for the SLT and OPT layouts, respectively. Overall, Variable Substitution demonstrates superior performance, particularly in the SLT representation, which appears to be more sensitive to structural changes.

Figure~\ref{fig:slt} illustrates the results for the SLT structure, which captures the spatial layout of formula symbols. In this context, Variable Substitution shows a distinct advantage. Under the ``full relevance'' setting, it achieves a top bpref score of 0.59, yielding a significant margin over the next best methods, which score at most 0.54. This significant gap underscores the importance of preserving the topological structure. Generic augmentations, such as Node Drop or Edge Drop, can severely disrupt the spatial arrangement (e.g., removing a superscript), thereby corrupting the formula's meaning. In contrast, Variable Substitution maintains the complete layout, enabling the model to learn the formula's abstract structure more effectively. Under the ``partial relevance'' threshold, it again achieves the highest score of 0.70, reinforcing its superiority.

A similar, though less pronounced, trend is observed for the OPT structure shown in Figure~\ref{fig:opt}, which represents the formula's operational hierarchy. Variable Substitution consistently outperforms other techniques across all batch sizes, achieving a bpref score of 0.58 in the ``full relevance'' setting, compared to 0.55 for the random augmentation. In the ``partial relevance'' setting, Variable Substitution and random strategy lead with a score of 0.70. It suggests that the operational semantics of OPTs may be slightly more resilient to random alterations than the strict spatial rules of SLTs. Nevertheless, the consistent lead of Variable Substitution confirms that preserving the integrity of the operator-operand tree is still the most effective strategy.

Across both graph representations, we note two general trends. First, larger batch sizes, which are typically expected to improve contrastive learning by providing more negative examples, only yield marginal performance gains. Second, the results are highly stable; we repeated each experiment 5 times, and the standard deviations were minimal across all settings (typically 0.001 to 0.009). These findings collectively underscore the effectiveness and robustness of Variable Substitution as a structure-preserving augmentation technique for math formula search.

%% file: contents/disc.tex
\section{Discussion}

This paper introduces a simple yet effective augmentation technique, Variable Substitution, for graph contrastive learning in the context of math formula search. Through extensive experiments, we demonstrate that this domain-specific method outperforms generic augmentation strategies, particularly in identifying structurally similar formulas. Our results suggest that preserving the core structural relationships between symbols and variables is critical to improving formula retrieval performance.

Future research could explore more sophisticated or targeted augmentation techniques that preserve mathematical semantics while increasing the diversity of the training data. Additionally, we are interested in applying this structure-preserving augmentation approach to other IR tasks involving structured data, such as chemical formula retrieval.